\documentclass[aps,onecolumn]{revtex4-2}
\usepackage{graphicx}
\usepackage{epsfig}
\usepackage{epstopdf}
\usepackage{amsfonts}
\usepackage{amssymb}
\usepackage{amsbsy}
\usepackage{amsmath}
\usepackage{mathrsfs}
\usepackage{latexsym}
\usepackage{natbib}
\usepackage{bm}
\usepackage{color}
\usepackage[a4paper, total={6.5in, 10.1in}]{geometry}
\DeclareMathOperator{\sech}{sech}

\usepackage{xifthen}
\newcommand{\dalembert}[1][]{\ifthenelse{\isempty{#1}}{\Box}{#1\Box}}

\usepackage{braket}
\usepackage{slashed}
\usepackage{pgfplots}
\numberwithin{equation}{section}

\usepackage{tikz}
\usetikzlibrary{shapes,arrows,shadows}

\begin{document}

\title{Ward identities under the frame transformations in curved spacetime}

\author{Susobhan Mandal}
\email{sm12ms085@gmail.com}

\affiliation{ Department of Physics, 
Indian Institute of Technology Bombay,\\
Mumbai 400076, India }


\begin{abstract}
\begin{center}
\underline{\textbf{Abstract}}
\end{center}
Scalar-tensor theories of gravity are considered to be competitors to Einstein’s theory of general relativity for the description of classical gravity, as they are used to build feasible models for cosmic inflation. These theories can be formulated both in the Jordan and Einstein frame, which are related by a Weyl transformation with a field transformation, known together as a frame transformation. These theories formulated in the above two frames are often considered to be equivalent from the point of view of classical theory. However, this is no longer true from the quantum field theoretical perspective. In the present article, we show that the Ward identities derived in the above two frames are not connected through the frame transformation. This shows that the quantum field theories formulated in these two frames are not equivalent to each other. Moreover, this inequivalence is also shown by comparing the effective actions derived in these two frames.


\end{abstract}

\maketitle

\section{Introduction}
Einstein's theory of general relativity becomes successful in describing the classical theory of gravity. However, there are other alternate classes of theories of gravity that are not completely eliminated through experimental observations. One such class of theories is the scalar-tensor theories of gravity \cite{wagoner1970scalar, barker1978general, ross1972scalar}. The scalar-tensor theories of gravity are often used as a credible model for our expanding Universe, and it also becomes successful in predicting cosmic inflation through the Higgs mechanism \cite{fakir1990induced, fakir1990improvement, bezrukov2009standard}. Further, cosmic inflation with different kinds of non-minimal couplings between Ricci scalar and scalar field has also been studied in \cite{hertzberg2010inflation}. Moreover, these theories have also been tested with the cosmological observations in \cite{campista2017testing, oda2018nonminimal} and the couplings within these theories are tightly constrained. In order to understand the nature of a particle physics model that underlies the inflationary scenario of the early Universe, these models with non-minimal coupling between gravity and scalar field have also been studied within the standard model of particle physics in \cite{barvinsky2008inflation, bezrukov2009standard, bostan2019quartic}.

The non-minimal couplings between gravity and scalar fields in the scalar-tensor theories of gravity make the computation of different physical observables difficult. This is described as a theory in the Jordan frame. However, there exists a set of transformations \cite{flanagan2004conformal} that leads to a theory with new metric and redefined scalar fields without any non-minimal couplings. This is described as the same theory in the Einstein frame, as it mimics the structure of the Einstein-Hilbert action in the general theory of relativity. In many models \cite{chiba2013conformal, catena2007einstein, ohta2018quantum, chakraborty2019dynamical, pandey2017equivalence}, it has been shown that the physical descriptions are frame independent, hence, the Jordan and Einstein frame are equivalent based on some observations in these models. Moreover, this claim is often made at the level of quantum theories. On the other hand, in several works \cite{kamenshchik2015question, faraoni1999einstein, jarv2007scalar, macias2001jordan, herrero2016anomalies, karam2017frame}, it has also been pointed out that the above two frames are inequivalent both at the quantum and classical level \cite{capozziello2010physical, kamenshchik2016transformations} in many field theories. In particular, the inequivalence between these two frames is shown in \cite{capozziello2010physical, kamenshchik2016transformations} at the classical level by comparing different observables and by looking at the crossing of singularities in cosmology. Further, it is also shown in many models that the null energy condition is violated in one of the frames despite they are related through the Weyl transformation. Therefore, whether these two frames are equivalent or not is still a debatable issue.

Symmetries in field theories play a significant role in understanding the dynamics of the corresponding system. Most importantly, corresponding to each continuous symmetry, there exists a conserved Noether current and its corresponding Noether charge. Moreover, the symmetries of a quantum field theory impose certain constraints among the correlation functions, known as the Ward identities \cite{ward1950identity, takahashi1957generalized}. These identities play a crucial role in gauge theories for renormalizing these theories \cite{taylor1971ward, slavnov1972ward, kluberg1975ward} and they are also useful in determining the scattering amplitudes \cite{low2018ward}. In order to address the question of equivalence between the Jordan and Einstein frame, we derive the Ward identities in a quantum field theory from their respective frames and compare them. We show explicitly that the Ward identities from an observer in the Einstein frame are not the same as the Ward identities from a different observer in the Jordan frame, related by the frame transformations. The charge conservation equations in both the frames are the same at the classical level, however, this is not true at the quantum level, which is shown using the functional integral approach. This leads to the conclusion that the above two frames are not equivalent.

\section{Frame transformations in classical field theory in curved spacetime}
In this section, the frame transformation between the Jordan and Einstein frame is briefly reviewed \cite{nozari2009comparison, morris2014consistency} for a class of scalar-tensor theories of gravity as a preliminary material for our later studies. We consider a class of scalar field theories in curved spacetime described by the following action
\begin{equation}\label{Jordan action}
\mathcal{S}_{g}[\phi]=\int\sqrt{-g}d^{4}x\Big[\mathcal{F}(\phi)R-\frac{1}{2}g^{\mu\nu}\nabla_{\mu}\phi\nabla_{\nu}\phi-V(\phi)\Big],
\end{equation}
where $R$ is the Ricci scalar, $V(\phi)$ is the potential for field $\phi$, and $\mathcal{F}(\phi)$ describes a non-minimal coupling between Ricci scalar and the field $\phi$. The index $g$ in the action is used as the above action in the Jordan frame is defined \textit{w.r.t} the metric $g_{\mu\nu}$. The minimization of the above action \textit{w.r.t} the metric $g_{\mu\nu}$ gives the following field equation
\begin{equation}\label{field equation Jordan}
\mathcal{F}(\phi)G_{\mu\nu}=\frac{1}{2}T_{\mu\nu}-g_{\mu\nu}\Box\mathcal{F}(\phi)+\nabla_{\mu}\nabla_{\nu}\mathcal{F}(\phi),
\end{equation}
where $G_{\mu\nu}=R_{\mu\nu}-\frac{1}{2}g_{\mu\nu}R$ is the Einstein tensor and $T_{\mu\nu}$ is the stress-energy tensor of the scalar field given by
\begin{equation}
T_{\mu\nu}=\nabla_{\mu}\phi\nabla_{\nu}\phi-\frac{1}{2}g_{\mu\nu}\nabla_{\alpha}\phi\nabla^{\alpha}\phi-g_{\mu\nu}V(\phi).
\end{equation}
On the other hand, the variation \textit{w.r.t} the field $\phi$ provides the Klein-Gordon equation
\begin{equation}
\Box\phi+R\mathcal{F}_{\phi}(\phi)-V_{\phi}(\phi)=0,
\end{equation}
where $\mathcal{F}_{\phi}(\phi)=\frac{d\mathcal{F}}{d\phi}$ and $V_{\phi}(\phi)=\frac{dV}{d\phi}$. In order to write the action (\ref{Jordan action}) in the Einstein frame, we make the following Weyl transformation from the Jordan frame
\begin{equation}
\tilde{g}_{\mu\nu}=e^{2\omega}g_{\mu\nu},
\end{equation}
under which the Lagrangian density of the action (\ref{Jordan action}) becomes
\begin{align}\label{relation 1}
\sqrt{-g}\Big[\mathcal{F}(\phi)R-\frac{1}{2}g^{\mu\nu}\nabla_{\mu}\phi\nabla_{\nu}\phi-V(\phi)\Big] & =\sqrt{-\tilde{g}}e^{-2\omega}\Big[\mathcal{F}(\phi)\tilde{R}-6\mathcal{F}(\phi)\overset{\sim}{\Box}\omega-6\mathcal{F}(\phi)\tilde{\nabla}_{\alpha}\omega\tilde{\nabla}^{\alpha}\omega \nonumber\\
 & -\frac{1}{2}\tilde{g}^{\mu\nu}\tilde{\nabla}_{\mu}\phi\tilde{\nabla}_{\nu}\phi-e^{-2\omega}V(\phi)\Big],
\end{align}
where $\tilde{R}$ and $\tilde{\nabla}_{\mu}$ are the Ricci scalar and covariant derivatives, respectively defined \textit{w.r.t} the metric $\tilde{g}_{\mu\nu}$. In order to define this theory in the Einstein frame \textit{w.r.t} the metric $\tilde{g}_{\mu\nu}$, we impose the following condition
\begin{equation}
e^{2\omega}=2\mathcal{F}.
\end{equation}
This is consistent since $\mathcal{F}$ must be positive definite. Using the above condition in the equation (\ref{relation 1}), we obtain the following relation
\begin{align}\label{relation 2}
\sqrt{-g} & \Big[\mathcal{F}(\phi)R-\frac{1}{2}g^{\mu\nu}\nabla_{\mu}\phi\nabla_{\nu}\phi-V(\phi)\Big]\\
=\sqrt{-\tilde{g}} & \Big[\frac{1}{2}\tilde{R}-3\overset{\sim}{\Box}\omega-\frac{3\mathcal{F}_{\phi}^{2}(\phi)+\mathcal{F}(\phi)}{4\mathcal{F}^{2}(\phi)}\tilde{\nabla}_{\alpha}\phi\tilde{\nabla}^{\alpha}\phi-\frac{V(\phi)}{4\mathcal{F}^{2}}\Big]. \nonumber
\end{align}
In terms of redefined field variables and the field potential satisfying the following relations
\begin{equation}\label{field redefinition}
d\tilde{\phi}=\sqrt{\frac{3\mathcal{F}_{\phi}^{2}(\phi)+\mathcal{F}(\phi)}{4\mathcal{F}^{2}(\phi)}}d\phi, \ \tilde{V}(\tilde{\phi})=\frac{V(\phi)}{4\mathcal{F}^{2}(\phi)},
\end{equation} 
the \textit{r.h.s} of (\ref{relation 2}) can be expressed as
\begin{align}
\sqrt{-g}\Big[\mathcal{F}(\phi)R-\frac{1}{2}g^{\mu\nu}\nabla_{\mu}\phi\nabla_{\nu}\phi-V(\phi)\Big] & =\sqrt{-\tilde{g}}\Big[\frac{1}{2}\tilde{R}-\frac{1}{2}\tilde{\phi}_{;\alpha}\tilde{\phi}_{;}^{ \ \alpha}-\tilde{V}(\tilde{\phi})\Big]+\text{boundary term}, 
\end{align}
where the boundary term is $\sqrt{-\tilde{g}}\overset{\sim}{\Box}\omega$ which can be avoided by choosing suitable boundary conditions on the field $\phi$. Further, the field equation (\ref{field equation Jordan}) in the Jordan frame reduces to the following equation
\begin{equation}
\tilde{G}_{\mu\nu}=\tilde{\phi}_{;\mu}\tilde{\phi}_{;\nu}-\frac{1}{2}\tilde{g}_{\mu\nu}\tilde{\phi}_{;\alpha}\tilde{\phi}_{;}^{ \ \alpha}+\tilde{g}_{\mu\nu}\tilde{V}(\tilde{\phi}),
\end{equation}
using the Weyl transformation and the field redefinition (\ref{field redefinition}). This is nothing but the Einstein's field equations in general relativity.

\section{Symmetries in quantum field theories}
In this section, we briefly discuss the construction of Noether's current and its corresponding conserved charge in classical field theory. Further, the Ward identities in quantum field theory are also briefly reviewed \cite{Peskin:1995ev} as a preliminary material for our main discussion in the next section.

\subsection{Noether's theorem and conserved charge}
According to Noether's theorem, every local continuous symmetry of a classical field theory corresponds to a conserved charge. Let us consider a classical field theory with action $\mathcal{A}[\phi]$ which is invariant under the following infinitesimal transformation
\begin{equation}\label{transformation}
\delta_{\epsilon}\phi(x)=\epsilon f(\phi,\partial\phi),
\end{equation} 
where $\epsilon$ is an infinitesimal small parameter. Since the action $\mathcal{A}[\phi]$ is invariant under the global transformation, we expect the variation of action under the local transformation to be of the following form
\begin{equation}\label{transform action}
\delta_{\epsilon}\mathcal{A}[\phi]=-\int_{\mathcal{M}}\sqrt{-g}d^{4}x \ j^{\mu}(x)\partial_{\mu}\epsilon(x),
\end{equation}
where $(\mathcal{M},g)$ is the underlying spacetime manifold. In the case of both local or global invariance, the above equation is equivalent to the following expression
\begin{equation}
\delta_{\epsilon}\mathcal{A}[\phi]=\int_{\mathcal{M}}\sqrt{-g}d^{4}x \ \epsilon\nabla_{\mu}j^{\mu}.
\end{equation}
However, on the support of the equation of motion, $\delta_{\epsilon}\mathcal{A}[\phi]=0$ even under the local transformation (\ref{transformation}) where $\epsilon$ is spacetime coordinate dependent function. As a consequence, we obtain covariant conservation of the 4-current $j^{\mu}$ 
\begin{equation}\label{conservation}
\nabla_{\mu}j^{\mu}=\frac{1}{\sqrt{-g}}\partial_{\mu}(\sqrt{-g}g^{\mu\nu}j_{\nu})=0,
\end{equation} 
whenever the equations of motion hold. The corresponding charge over a co-dimension one hypersurface $\mathcal{N}$ is defined as
\begin{equation}
\mathcal{Q}[\mathcal{N}]\equiv\int_{\mathcal{N}}\sqrt{-g}d^{3}x \ j_{\mu}(x)n^{\mu}(x),
\end{equation}
where $n^{\mu}$ is the normal vector on the hypersurface $\mathcal{N}$. Using the Stoke's theorem and the relation (\ref{conservation}), it can be shown that $\mathcal{Q}[\mathcal{N}_{0}]=\mathcal{Q}[\mathcal{N}_{1}]$ where $\mathcal{N}_{0,1}$ are the two co-dimension one hypersurfaces bounding a region $\mathcal{M}'\subset\mathcal{M}$. If $\mathcal{N}_{0,1}$ are chosen to be $t=\text{constant}$ hypersurfaces, then this shows that the charge $\mathcal{Q}$ is conserved under time evolution. This is the classical version of Noether's theorem.

\subsection{Ward identities}
The quantum version of Noether's theorem or in other words, consequences of symmetries at the quantum level can be derived in terms of the Ward identities. The functional integral approach is one of the ways in which a classical field theory can be quantized. The generating functional is given by
\begin{equation}
\mathcal{Z}[J]=\int\mathcal{D}\Phi \ e^{i\mathcal{A}[\Phi]+i\int\sqrt{-g}d^{4}x \ J(x)\Phi(x)},
\end{equation}
which generates correlation functions in terms of functional derivatives \textit{w.r.t} source field $J(x)$. $\Phi$ denotes a collection of scalar fields and $J$ denotes a collection of corresponding conjugate sources. The action $\mathcal{A}[\Phi]$ is invariant under the field transformation $\Phi\rightarrow\Phi'=\Phi+\delta_{\epsilon}\Phi$. As a consequence, we obtain the following relation
\begin{align}
\mathcal{Z} & [J=0]=\int\mathcal{D}\Phi \ e^{i\mathcal{A}[\Phi]}=\int\mathcal{D}\Phi' \ e^{i\mathcal{A}[\Phi']}\\
 & =\int\mathcal{D}\Phi \ e^{i\mathcal{A}[\Phi]}\Bigg[1-i\int_{\mathcal{M}}\sqrt{-g}d^{4}x \ g^{\mu\nu}(x)j_{\mu}(x)\partial_{\nu}\epsilon(x)\Bigg] \nonumber,
\end{align} 
in lowest order where we have used parametrization invariance of $\mathcal{Z}[J=0]$ and the equation (\ref{transform action}). We also assume that the functional measure is invariant under the transformation $\Phi\rightarrow\Phi'=\Phi+\delta_{\epsilon}\Phi$ which is not true always \cite{fujikawa2003topological, fujikawa1980path}. The above relation implies
\begin{equation}
\int_{\mathcal{M}}\sqrt{-g}d^{4}x \ \epsilon(x)\nabla_{\mu}\langle j^{\mu}(x)\rangle=0,
\end{equation}
which is nothing but the quantum version of conservation equation (\ref{conservation}), where
\begin{equation}
\langle j^{\mu}(x)\rangle=\int\mathcal{D}\Phi \ e^{i\mathcal{A}[\Phi]} j^{\mu}(x).
\end{equation}
Given a set of operators $\{\hat{\mathcal{O}}_{i}(\Phi(x_{i}))\}$, their correlation functions are defined as
\begin{align}
\langle\hat{\mathcal{O}}_{1}(\Phi(x_{1}))\ldots\hat{\mathcal{O}}_{n}(\Phi(x_{n}))\rangle =\int\mathcal{D}\Phi \ \mathcal{O}_{1}(\Phi(x_{1}))\ldots\mathcal{O}_{n}(\Phi(x_{n})) \ e^{i\mathcal{A}[\Phi]}.
\end{align}
The infinitesimal change in the functions $\mathcal{O}_{i}(\Phi(x_{i}))$ under the infinitesimal symmetry transformation $\Phi\rightarrow\Phi'=\Phi+\delta_{\epsilon}\Phi$  is given by
\begin{equation}
\mathcal{O}_{i}(\Phi(x_{i}))\mapsto\mathcal{O}_{i}(\Phi(x_{i})+\delta_{\epsilon}\Phi(x_{i}))=\mathcal{O}_{i}(\Phi(x_{i}))+\epsilon(x_{i})\delta\mathcal{O}_{i}(x_{i}).
\end{equation}
Considering both the changes of action and the above set of operators, we obtain the following relation
\begin{equation}
\begin{split}
\int\mathcal{D}\Phi \ & e^{i\mathcal{A}[\Phi]}\prod_{i=1}^{n}\mathcal{O}_{i}(\Phi(x_{i}))=\int\mathcal{D}\Phi' \ e^{i\mathcal{A}[\Phi']}\prod_{i=1}^{n}\mathcal{O}_{i}(\Phi'(x_{i}))\\
=\int\mathcal{D} & \Phi \ e^{i\mathcal{A}[\Phi]}\Bigg[1-i\int_{\mathcal{M}}\sqrt{-g}d^{4}x \ g^{\mu\nu}(x)j_{\mu}(x)\partial_{\nu}\epsilon(x)\Bigg]\\
\times\Bigg[ & \prod_{i=1}^{n}\mathcal{O}_{i}(\Phi(x_{i}))+\sum_{i=1}^{n}\epsilon(x_{i})\delta\mathcal{O}_{i}(x_{i})\prod_{j\neq i}\mathcal{O}_{j}(\Phi(x_{j}))\Bigg].
\end{split}
\end{equation}
Therefore, we obtain the following relation in the first order in $\epsilon$
\begin{equation}\label{identity 1}
i\int_{\mathcal{M}}\sqrt{-g}d^{4}x \ \epsilon(x)\nabla_{\mu}\langle j^{\mu}(x)\prod_{i=1}^{n}\mathcal{O}_{i}(\Phi(x_{i}))\rangle =-\sum_{i=1}^{n}\epsilon(x_{i})\langle\delta\mathcal{O}_{i}(x_{i})\prod_{j\neq i}\mathcal{O}_{j}(\Phi(x_{j}))\rangle,
\end{equation} 
where an integration by parts is done. Using the following identity
\begin{equation}
\epsilon(x_{i})\delta\mathcal{O}_{i}(x_{i})=\int_{\mathcal{M}}\delta^{(4)}(x-x_{i})\epsilon(x)\delta\mathcal{O}_{i}(x)d^{4}x,
\end{equation} 
equation (\ref{identity 1}) reduces to the following constraint between the correlation functions
\begin{equation}\label{Ward identities}
i\nabla_{\mu}\langle j^{\mu}(x)\prod_{i=1}^{n}\mathcal{O}_{i}(\Phi(x_{i}))\rangle =-\sum_{i=1}^{n}\frac{\delta^{(4)}(x-x_{i})}{\sqrt{-g(x)}}\langle\delta\mathcal{O}_{i}(x_{i})\prod_{j\neq i}\mathcal{O}_{j}(\Phi(x_{j}))\rangle,
\end{equation}
known as the Ward identities. If we consider a region $\mathcal{M}'\subset\mathcal{M}$ containing a collection of points $\{x_{j}\}_{j\in I}$ bounded by the hypersurfaces $\mathcal{N}_{1,2}$ where $I\subset\{1,2,\ldots,n\}$ is an index set, then the Ward identities (\ref{Ward identities}) give rise to the following relations
\begin{equation}
\langle\mathcal{Q}[\mathcal{N}_{2}]\prod_{i=1}^{n}\mathcal{O}_{i}(\Phi(x_{i}))\rangle-\langle\mathcal{Q}[\mathcal{N}_{1}]\prod_{i=1}^{n}\mathcal{O}_{i}(\Phi(x_{i}))\rangle=i\sum_{i\in I}\langle\delta\mathcal{O}_{i}(x_{i})\prod_{j\neq i}\mathcal{O}_{j}(\Phi(x_{j}))\rangle.
\end{equation}
On the other hand, if $\mathcal{M}'=\mathcal{M}$ is a closed manifold (compact without boundary), then the above relation reduces to
\begin{equation}
\sum_{i=1}^{n}\langle\delta\mathcal{O}_{i}(x_{i})\prod_{j\neq i}\mathcal{O}_{j}(\Phi(x_{j}))\rangle=0.
\end{equation}
The above results also hold for a global symmetry.

\section{Frame transformations at the quantum level}
\subsection{Ward identities due to an internal symmetry}
In order to show that the Jordan frame and Einstein frame of a field theory are inequivalent at the quantum level, we consider a specific example of a complex scalar field theory invariant under the global $U(1)$ transformation. Within this model, we show that the Ward identities in these two frames are not equivalent to each other. In order to show this, we find the Ward identities from the Einstein frame and compare them with the similar identities from the Jordan frame but in terms of field variables used in the Einstein frame. Now onward, we use $\sim$ to denote the quantities in the Einstein frame.

We consider the following action 
\begin{equation}
S_{J}[\phi,\phi^{\dagger};g]=\int\sqrt{-g}d^{4}x\Big[\left(\frac{1}{16\pi G}-\xi\phi^{\dagger}\phi\right)R-g^{\mu\nu}\nabla_{\mu}\phi^{\dagger}\nabla_{\nu}\phi-V(\phi^{\dagger}\phi)\Big],
\end{equation}    
where the Newton's gravitational constant $G$ is written explicitly and the index $J$ is used to indicate that the above action is defined in the Jordan frame. For mathematical simplicity, we write the above action with conformal coupling $\xi=\frac{1}{6}$ in the polar parametrization $\phi=\frac{1}{\sqrt{2}}\rho e^{i\theta}, \ \phi^{\dagger}=\frac{1}{\sqrt{2}}\rho e^{-i\theta}$
\begin{align}
S_{J}[\rho,\theta;g] & =\int\sqrt{-g}d^{4}x\Big[\left(\frac{1}{16\pi G}-\frac{1}{12}\rho^{2}\right)R-\frac{1}{2}g^{\mu\nu}\nabla_{\mu}\rho\nabla_{\nu}\rho-\frac{1}{2}g^{\mu\nu}\rho^{2}\nabla_{\mu}\theta\nabla_{\nu}\theta-V(\rho)\Big].\nonumber
\end{align}
As the action suggests, the conserved current is given by $j^{\mu}=\rho^{2}\nabla^{\mu}\theta$. As we discussed earlier, doing the following Weyl and field transformations
\begin{align}\label{frame transform}
\tilde{g}_{\mu\nu} & =e^{2\omega}g_{\mu\nu}, \ e^{2\omega}=1-\frac{4\pi G}{3}\rho^{2} \nonumber\\
d\tilde{\rho} & =\frac{1}{1-\frac{4\pi G}{3}\rho^{2}}d\rho\implies\rho=\sqrt{\frac{3}{4\pi G}}\tanh\left(\sqrt{\frac{4\pi G}{3}}\tilde{\rho}\right) \nonumber\\ 
\tilde{V} & (\tilde{\rho}) =e^{-4\omega}V(\rho),
\end{align}
give rise to the following action in the Einstein frame
\begin{align}
S_{E} & [\tilde{\rho},\tilde{\theta};\tilde{g}]=\int\sqrt{-\tilde{g}}d^{4}x\Big[\frac{1}{16\pi G}R-\frac{1}{2}\tilde{g}^{\mu\nu}\tilde{\nabla}_{\mu}\tilde{\rho}\tilde{\nabla}_{\nu}\tilde{\rho}-\frac{3}{8\pi G}\tilde{g}^{\mu\nu}\sinh^{2}\left(\sqrt{\frac{4\pi G}{3}}\tilde{\rho}\right)\tilde{\nabla}_{\mu}\theta\tilde{\nabla}_{\nu}\theta-\tilde{V}(\tilde{\rho})\Big],
\end{align}
up to the boundary terms which are not important here. We also assumed that $\Big|\sqrt{\frac{4\pi G}{3}}\rho\Big|\leq1$. The conserved current due to the  $U(1)$ symmetry in the Einstein frame is given by $\tilde{j}^{\mu}=\frac{3}{4\pi G}\sinh^{2}\left(\sqrt{\frac{4\pi G}{3}}\tilde{\rho}\right)\tilde{\nabla}^{\mu}\theta$ which follows from the equations of motion in the Einstein frame. This can also be obtained from the frame transformation (\ref{frame transform}) of the covariant conservation equation of the conserved current in the Jordan frame. Hence, at the classical level, these theories in the Jordan and Einstein frames are equivalent.

Now we write the Ward identities in both the frame through the same set of correlation functions in order to compare them. First, we consider the Jordan frame in which the generating functional is given by
\begin{align}\label{generating functional Jordan}
\mathcal{Z}[J_{\rho},J_{\theta}] & =\int\mathcal{D}\rho\bar{\mathcal{D}}_{\rho}\theta \ e^{iS_{J}[\rho,\theta;g]+i\int\sqrt{-g}d^{4}x[J_{\rho}\rho+J_{\theta}\theta]}, \ 
\bar{\mathcal{D}}_{\rho}\theta=\prod_{x}[\rho(x)d\theta(x)].
\end{align}
Considering the set of observables $\{\mathcal{O}_{i}(\rho(x_{i}),\theta(x_{i}),g(x_{i}))\}$, the Ward identities in the Jordan frame are expressed as
\begin{align}\label{Ward identity Jordan}
i & \nabla_{\mu}\langle j^{\mu}(x)\prod_{i=1}^{n}\mathcal{O}_{i}(\rho(x_{i}),\theta(x_{i}),g(x_{i}))\rangle_{J} \nonumber\\
= & -\sum_{i=1}^{n}\frac{\delta^{(4)}(x-x_{i})}{\sqrt{-g(x)}}\langle\delta\mathcal{O}_{i}(x_{i})\prod_{j\neq i}\mathcal{O}_{j}(\rho(x_{i}),\theta(x_{i}),g(x_{i}))\rangle_{J},
\end{align}
where $\epsilon\delta\mathcal{O}_{i}(x_{i})$ is the change in $\mathcal{O}_{i}(\rho(x_{i}),\theta(x_{i}),g(x_{i}))$ under the infinitesimal transformation $\theta(x)\rightarrow\theta'(x)=\theta(x)+\epsilon$ and
\begin{align}
\langle\prod_{i=1}^{n}\mathcal{O}_{i}(\rho(x_{i}),\theta(x_{i}),g(x_{i}))\rangle_{J}=\int\mathcal{D}\rho\bar{\mathcal{D}}_{\rho}\theta \ \prod_{i=1}^{n}\mathcal{O}_{i}(\rho(x_{i}),\theta(x_{i}),g(x_{i})) \ e^{iS_{J}[\rho,\theta;g]}.
\end{align}
Similarly, we also obtain $\nabla_{\mu}\langle j^{\mu}(x)\rangle=0$. Using only the naive Weyl and field transformations, the equations in (\ref{Ward identity Jordan}) are expected to become
\begin{align}\label{naive}
i & \tilde{\nabla}_{\mu}\langle \tilde{j}^{\mu}(x)\prod_{i=1}^{n}\tilde{\mathcal{O}}_{i}(\tilde{\rho}(x_{i}),\theta(x_{i}),\tilde{g}(x_{i}))\rangle_{E} \nonumber\\
= & -\sum_{i=1}^{n}\Big[\frac{\delta^{(4)}(x-x_{i})}{\sqrt{-\tilde{g}(x)}}\langle\delta\tilde{\mathcal{O}}_{i}(x_{i})\prod_{j\neq i}\tilde{\mathcal{O}}_{j}(\tilde{\rho}(x_{i}),\theta(x_{i}),\tilde{g}(x_{i}))\rangle_{E}\Big] ,
\end{align}
according to an observer in the Einstein frame where $\tilde{\mathcal{O}}_{i}(\tilde{\rho}(x_{i}),\theta(x_{i}),\tilde{g}(x_{i}))=\mathcal{O}_{i}(\rho(x_{i}),\theta(x_{i}),g(x_{i}))$, and 
\begin{align}
\langle & \prod_{i=1}^{n}\tilde{\mathcal{O}}_{i}(\tilde{\rho}(x_{i}),\theta(x_{i}),\tilde{g}(x_{i}))\rangle_{E}=\int\mathcal{D}\tilde{\rho}\bar{\mathcal{D}}_{\tilde{\rho}}\theta \ \Bigg[\prod_{i=1}^{n}\tilde{\mathcal{O}}_{i}(\tilde{\rho}(x_{i}),\theta(x_{i}),\tilde{g}(x_{i}))\Bigg] \ e^{iS_{E}[\tilde{\rho},\theta;\tilde{g}]}.
\end{align} 
However, the above relations in (\ref{naive}) are not the Ward identities in Einstein frame from the point of view of an observer in the Jordan frame, which is shown below. 

In order to define the Ward identities in the Einstein frame from the point of view of an observer in the Jordan frame, the Weyl transformation of weight in the measure of functional integral must be taken into account following the appendix. Here, we use a regularization, in which the functional integral measure defined in (\ref{measure}) changes by the following factor from the Jordan to Einstein frame
\begin{align}
e^{S_{1}[\tilde{\rho}; \tilde{g}]}\equiv\prod_{x}e^{-2\omega[\tilde{\rho}(x)]}=e^{-2\text{Tr}[\omega[\tilde{\rho}(\hat{x})]]} = e^{2\int\sqrt{-\tilde{g}(x)}d^{4}x\log\cosh\left(\sqrt{\frac{4\pi G}{3}}\tilde{\rho}(x)\right)}.
\end{align}

As a result, from the point of view of an observer in the Jordan frame, the expression of generating functional in (\ref{generating functional Jordan}) without the sources in the Einstein frame becomes
\begin{align}
\mathcal{Z} & [J_{\rho}=0,J_{\theta}=0]=\int\mathcal{D}\rho\bar{\mathcal{D}}_{\rho}\theta \ e^{iS_{J}[\rho,\theta;g]}=\int\mathcal{D}\rho\bar{\mathcal{D}}_{\rho}\theta \ e^{iS_{J}[\rho[\tilde{\rho}],\theta;g[\tilde{g}]]} \nonumber\\
 & =\int\mathcal{D}\tilde{\rho}\bar{\mathcal{D}}_{\tilde{\rho}}\theta \ \Big[\text{det}[\mathcal{J}[\tilde{\rho}]]e^{iS_{E}[\tilde{\rho},\theta;\tilde{g}]+S_{1}[\tilde{\rho}; \tilde{g}]}\Big], \ \bar{\mathcal{D}}_{\tilde{\rho}}\theta=\prod_{x}[\tilde{\rho}(x)d\theta(x)],
\end{align}
where $\text{det}[\mathcal{J}[\tilde{\rho}]]$ is the determinant of the Jacobian of the field transformations. The expression of $\text{det}[\mathcal{J}[\tilde{\rho}]]$ in the Einstein frame is given by
\begin{equation}
\begin{split}
\mathcal{J}[\tilde{\rho}]=\frac{\delta\rho}{\delta\tilde{\rho}}\frac{\rho}{\tilde{\rho}} & =\frac{\sech^{2}\left(\sqrt{\frac{4\pi G}{3}}\tilde{\rho}\right)\tanh\left(\sqrt{\frac{4\pi G}{3}}\tilde{\rho}\right)}{\sqrt{\frac{4\pi G}{3}}\tilde{\rho}}
\\
\implies\text{det}[\mathcal{J}[\tilde{\rho}]] & =e^{\int\sqrt{-\tilde{g}(x)}d^{4}x \ \log\left(\mathcal{J}[\tilde{\rho}(x)]\right)}\equiv e^{\mathcal{S}_{E}[\tilde{\rho};\tilde{g}]}.
\end{split}
\end{equation}
Hence, we obtain the following expression
\begin{equation}
\mathcal{Z}[J_{\rho}=0,J_{\theta}=0]=\int\mathcal{D}\tilde{\rho}\bar{\mathcal{D}}_{\tilde{\rho}}\theta \ e^{iS_{E}[\tilde{\rho},\theta;\tilde{g}]+S_{2}[\tilde{\rho}; \tilde{g}]},
\end{equation}
in the Einstein frame from the point of view of the observer in the Jordan frame, where $S_{2}[\tilde{\rho}; \tilde{g}]=\mathcal{S}_{E}[\tilde{\rho};\tilde{g}]+S_{1}[\tilde{\rho}; \tilde{g}]$. As a result, the Ward identities in the Einstein frame become the following
\begin{align}
i & \tilde{\nabla}_{\mu}\langle \tilde{j}^{\mu}(x)\prod_{i=1}^{n}\tilde{\mathcal{O}}_{i}(\tilde{\rho}(x_{i}),\theta(x_{i}),\tilde{g}(x_{i}))\rangle_{J-E} \nonumber\\
= & -\sum_{i=1}^{n}\frac{\delta^{(4)}(x-x_{i})}{\sqrt{-\tilde{g}(x)}}\langle\delta\tilde{\mathcal{O}}_{i}(x_{i})\prod_{j\neq i}\tilde{\mathcal{O}}_{j}(\tilde{\rho}(x_{i}),\theta(x_{i}),\tilde{g}(x_{i}))\rangle_{J-E},
\end{align} 
according to the observer in the Jordan frame, where
\begin{align}
\langle\prod_{i=1}^{n}\tilde{\mathcal{O}}_{i}(\tilde{\rho}(x_{i}),\theta(x_{i}),\tilde{g}(x_{i}))\rangle_{J-E}\equiv\int\mathcal{D}\tilde{\rho}\bar{\mathcal{D}}_{\tilde{\rho}}\theta \ \Big[\prod_{i=1}^{n}\tilde{\mathcal{O}}_{i}(\tilde{\rho}(x_{i}),\theta(x_{i}),\tilde{g}(x_{i})) \ e^{iS_{E}[\tilde{\rho},\theta;\tilde{g}]+S_{2}[\tilde{\rho}; \tilde{g}]}\Big].
\end{align} 
The above clearly shows a non-trivial change in the Ward identities due to the frame transformation from the Jordan frame to Einstein frame. Moreover, we also obtain the following relation
\begin{equation}
\tilde{\nabla}_{\mu}\langle \tilde{j}^{\mu}(x)\rangle_{J-E}=0,
\end{equation}  
which is different from the conservation equation $\tilde{\nabla}_{\mu}\langle \tilde{j}^{\mu}(x)\rangle_{E}=0$ in the Einstein frame, where the frame transformations are used naively. Therefore, the physical descriptions of a quantum field theory in the Jordan and Einstein frame are not equivalent. A similar conclusion can also be drawn for a local $U(1)$ invariant field theory using the gauge fields.

\subsection{Ward identities due to the diffeomorphism invariance}
For the sake of simplicity in showing the Ward identities due to the diffeomorphism invariance, we consider the following action of a real scalar field theory in the Jordan frame
\begin{equation}\label{action 2 in diff inv}
\mathcal{A}_{J}[\phi,g]=\int\sqrt{-g}d^{4}x\Bigg[\left(\frac{1}{16\pi G}-\frac{1}{12}\phi^{2}\right)R-\frac{1}{2}g^{\mu\nu}\nabla_{\mu}\phi\nabla_{\nu}\phi-U(\phi)\Bigg].
\end{equation}
The corresponding stress-energy tensor is given by the following expression
\begin{equation}
\begin{split}
T_{\mu\nu} & = - \frac{2}{\sqrt{-g}}\frac{\delta\mathcal{A}_{J}}{\delta g^{\mu\nu}}\\
 & = \Big[\nabla_{\mu}\phi\nabla_{\nu}\phi-\frac{1}{2}g_{\mu\nu}\nabla_{\rho}\phi\nabla^{\rho}\phi-g_{\mu\nu}U(\phi)\\
& +\frac{1}{6}(g_{\mu\nu}\Box-\nabla_{\mu}\nabla_{\nu})\phi^{2}+\frac{1}{6}\phi^{2}G_{\mu\nu}\Big],
\end{split}
\end{equation}  
which is covariantly conserved supported by the equation of motion for the field $\phi$. After doing the following set of transformations
\begin{align}\label{frame transform 2}
\tilde{g}_{\mu\nu} & =e^{2\omega}g_{\mu\nu}, \ e^{2\omega}=1-\frac{4\pi G}{3}\phi^{2}\\
\phi & =\sqrt{\frac{3}{4\pi G}}\tanh\left(\sqrt{\frac{4\pi G}{3}}\tilde{\phi}\right), \ \tilde{U}(\tilde{\phi}) =e^{-4\omega}U(\phi) \nonumber,
\end{align}
we obtain the following action in the Einstein frame
\begin{equation}
\mathcal{A}_{E}[\tilde{\phi};\tilde{g}]=\int\sqrt{-\tilde{g}}d^{4}x\Big[\frac{1}{16\pi G}R-\frac{1}{2}\tilde{g}^{\mu\nu}\tilde{\nabla}_{\mu}\tilde{\phi}\tilde{\nabla}_{\nu}\tilde{\phi}-\tilde{U}(\tilde{\phi})\Big].
\end{equation}
In the Einstein frame, the corresponding stress-energy tensor is given by the following expression
\begin{equation}
\tilde{T}_{\mu\nu}=\Big[\tilde{\nabla}_{\mu}\tilde{\phi}\tilde{\nabla}_{\nu}\tilde{\phi}-\frac{1}{2}\tilde{g}_{\mu\nu}\tilde{\nabla}_{\rho}\tilde{\phi}\tilde{\nabla}^{\rho}\tilde{\phi}-\tilde{g}_{\mu\nu}\tilde{U}(\tilde{\phi})\Big],
\end{equation}
which is also covariantly conserved supported by the equation of motion for the field $\tilde{\phi}$ coming from variation of the action $\mathcal{A}_{E}[\tilde{\phi};\tilde{g}]$ \textit{w.r.t} $\tilde{\phi}$.

The above theory in both the references are invariant under the infinitesimal diffeomorphism $x^{\mu}\rightarrow x'^{\mu}=x^{\mu}+\xi^{\mu}(x)$. Under this transformation, metric changes by $\delta_{\xi}g_{\mu\nu}(x)=\nabla_{\mu}\xi_{\nu}(x)+\nabla_{\nu}\xi_{\mu}(x)$ and the field changes by $\delta_{\xi}\phi(x)=-\xi^{\mu}(x)\nabla_{\mu}\phi(x)$. Since the following quantity defined in Jordan frame 
\begin{equation}
\mathcal{Z}_{J}=\mathcal{N}\int\mathcal{D}\phi \ e^{i\mathcal{A}_{J}[\phi,g]},
\end{equation} 
is invariant under the infinitesimal diffeomorphism, it can be shown easily that $\nabla_{\mu}\langle T^{\mu\nu}(x)\rangle=0$. $\mathcal{N}$ is a normalization constant which is used in order to absorb irrelevant Jacobian of field transformation. Moreover, the correlation functions between the observables $\{\mathcal{O}_{i}(\phi(x_{i}),g(x_{i}))\}$ satisfy the following relation
\begin{align}\label{diff invariance of partition func}
\langle & \prod_{i=1}^{n}\mathcal{O}_{i}(\phi(x_{i}),g(x_{i}))\rangle_{J}=\mathcal{N}\int\mathcal{D}\phi \ \prod_{i=1}^{n}\mathcal{O}_{i}(\phi(x_{i}),g(x_{i})) \ e^{i\mathcal{A}_{J}[\phi,g]} \nonumber\\
 & =\mathcal{N}\int\mathcal{D}\phi' \ \prod_{i=1}^{n}\mathcal{O}_{i}(\phi'(x_{i}),g'(x_{i})) \ e^{i\mathcal{A}_{J}[\phi',g']}\\
 & =\mathcal{N}\int\mathcal{D}\phi \ \prod_{i=1}^{n}\mathcal{O}_{i}(\phi'(x_{i}),g'(x_{i})) \ e^{i\mathcal{A}_{J}[\phi,g]+i\delta_{\xi}\mathcal{A}_{J}[\phi,g]} \nonumber,
\end{align} 
where 
\begin{equation}\label{diff relation1}
\delta_{\xi}\mathcal{A}_{J}[\phi,g]=\int\sqrt{-g}d^{4}x \ T^{\mu\nu}(x)\nabla_{\mu}\xi_{\nu}-\int d^{4}x\frac{\delta\mathcal{A}_{J}[\phi,g]}{\delta\phi(x)}\xi^{\mu}\nabla_{\mu}\phi,
\end{equation}
and
\begin{equation}\label{diff relation2}
\begin{split}
\mathcal{O}_{i}(\phi'(x_{i}) & ,g'(x_{i}))=\mathcal{O}_{i}(\phi(x_{i}),g(x_{i}))+\delta_{\phi}\mathcal{O}_{i}(\phi(x_{i}),g(x_{i}))+\delta_{g}\mathcal{O}_{i}(\phi(x_{i}),g(x_{i}))\\
\delta_{\phi}\mathcal{O}_{i} & (\phi(x_{i}),g(x_{i}))=-\int d^{4}x\xi^{\mu}(x)\nabla_{\mu}\phi(x)\frac{\delta\mathcal{O}_{i}(\phi(x_{i}),g(x_{i}))}{\delta\phi(x)}\\
\delta_{g}\mathcal{O}_{i} & (\phi(x_{i}),g(x_{i}))=\int d^{4}x\frac{\delta\mathcal{O}_{i}(\phi(x_{i}),g(x_{i}))}{\delta g_{\mu\nu}(x)}\nabla_{\mu}\xi_{\nu}(x)\\
 & =-\int d^{4}x\xi_{\nu}(x)\nabla_{\mu}\left(\frac{\delta\mathcal{O}_{i}(\phi(x_{i}),g(x_{i}))}{\delta g_{\mu\nu}(x)}\right).
\end{split}
\end{equation}   
Therefore, inserting the relations (\ref{diff relation1}, \ref{diff relation2}) in the equation (\ref{diff invariance of partition func}), we obtain the following relation
\begin{align}
0 & =i\int\mathcal{D}\phi \ e^{i\mathcal{A}_{J}[\phi,g]}\Bigg[\int\sqrt{-g}d^{4}x \ T^{\mu\nu}(x)\nabla_{\mu}\xi_{\nu}(x) \nonumber\\ 
- & \int d^{4}x\xi^{\mu}(x)\nabla_{\mu}\phi(x)\frac{\delta\mathcal{A}_{J}[\phi,g]}{\delta\phi(x)}\Bigg]\left( \prod_{i=1}^{n}\mathcal{O}_{i}(\phi(x_{i}),g(x_{i}))\right) \nonumber\\
+ & \sum_{j=1}^{n}\int\mathcal{D}\phi \ \delta_{g}\mathcal{O}_{j}(\phi(x_{j}),g(x_{j}))\prod_{i\neq j}^{n}\mathcal{O}_{i}(\phi(x_{i}),g(x_{i}))e^{i\mathcal{A}_{J}[\phi,g]} \nonumber\\
+ & \sum_{j=1}^{n}\int\mathcal{D}\phi \ \delta_{\phi}\mathcal{O}_{j}(\phi(x_{j}),g(x_{j}))\prod_{i\neq j}^{n}\mathcal{O}_{i}(\phi(x_{i}),g(x_{i}))e^{i\mathcal{A}_{J}[\phi,g]}.
\end{align}
Hence, from the above relation, we obtain the following Ward identities
\begin{align}
i & \nabla_{\mu}\langle T^{\mu\nu}(x)\prod_{i=1}^{n}\mathcal{O}_{i}(\phi(x_{i}),g(x_{i}))\rangle_{J}=-\frac{i}{\sqrt{-g(x)}}\Big\langle \nabla^{\nu}\phi(x)\frac{\delta\mathcal{A}_{J}[\phi,g]}{\delta\phi(x)}\prod_{i=1}^{n}\mathcal{O}_{i}(\phi(x_{i}),g(x_{i}))\Big\rangle_{J} \nonumber\\
 & -\sum_{j=1}^{n}\frac{1}{\sqrt{-g(x)}}\Big\langle\nabla_{\mu}\left(\frac{\delta\mathcal{O}_{j}(\phi(x_{j}),g(x_{j}))}{\delta g_{\mu\nu}(x)}\right)\prod_{i\neq j}^{n}\mathcal{O}_{i}(\phi(x_{i}),g(x_{i}))\Big\rangle_{J} \nonumber\\
 & -\sum_{j=1}^{n}\frac{1}{\sqrt{-g(x)}}\Big\langle\nabla^{\nu}\phi(x)\frac{\delta\mathcal{O}_{j}(\phi(x_{j}),g(x_{j}))}{\delta\phi(x)}\prod_{i\neq j}\mathcal{O}_{i}(\phi(x_{i}),g(x_{i}))\Big\rangle_{J}.
\end{align}
This seems to suggest naively that for an observer in the Einstein frame, the Ward identities due to the diffeomorphism invariance become
\begin{align}
i & \tilde{\nabla}_{\mu}\langle \tilde{T}^{\mu\nu}(x)\prod_{i=1}^{n}\tilde{\mathcal{O}}_{i}(\tilde{\phi}(x_{i}),\tilde{g}(x_{i}))\rangle_{E}=-\frac{i}{\sqrt{-\tilde{g}(x)}}\Big\langle\tilde{\nabla}^{\nu}\tilde{\phi}(x)\frac{\delta\mathcal{A}_{E}[\tilde{\phi},\tilde{g}]}{\delta\tilde{\phi}(x)}\prod_{i=1}^{n}\tilde{\mathcal{O}}_{i}(\tilde{\phi}(x_{i}),\tilde{g}(x_{i}))\Big\rangle_{E} \nonumber\\
 & -\sum_{j=1}^{n}\frac{1}{\sqrt{-\tilde{g}(x)}}\Big\langle\tilde{\nabla}_{\mu}\left(\frac{\delta\tilde{\mathcal{O}}_{j}(\tilde{\phi}(x_{j}),\tilde{g}(x_{j}))}{\delta\tilde{g}_{\mu\nu}(x)}\right)\prod_{i\neq j}^{n}\tilde{\mathcal{O}}_{i}(\tilde{\phi}(x_{i}),\tilde{g}(x_{i}))\Big\rangle_{E} \nonumber\\
 & -\sum_{j=1}^{n}\frac{1}{\sqrt{-\tilde{g}(x)}}\Big\langle\tilde{\nabla}^{\nu}\tilde{\phi}(x)\frac{\delta\tilde{\mathcal{O}}_{j}(\tilde{\phi}(x_{j}),\tilde{g}(x_{j}))}{\delta\tilde{\phi}(x)}\prod_{i\neq j}\tilde{\mathcal{O}}_{i}(\tilde{\phi}(x_{i}),\tilde{g}(x_{i}))\Big\rangle_{E},
\end{align}
where $\tilde{\mathcal{O}}_{i}(\tilde{\phi}(x_{i}),\tilde{g}(x_{i}))=\mathcal{O}_{i}(\phi(x_{i}),g(x_{i}))$, and 
\begin{equation}
\langle\tilde{\mathcal{O}}_{i}(\tilde{\phi}(x_{i}),\tilde{g}(x_{i}))\rangle_{E}=\int\mathcal{D}\tilde{\phi} \ \prod_{i=1}^{n}\tilde{\mathcal{O}}_{i}(\tilde{\phi}(x_{i}),\tilde{g}(x_{i})) e^{i\mathcal{A}_{E}[\tilde{\phi},\tilde{g}]}.
\end{equation} 
However, that is not the case from the point of view of an observer in the Jordan frame which we show now. Like earlier, it can be shown that the generating functional from the point of an observer in the Jordan frame without the sources can be expressed as
\begin{align}\label{GF E-J}
\mathcal{Z}_{J} & =\mathcal{N}\int\mathcal{D}\phi \ e^{i\mathcal{A}_{J}[\phi,g]}=\mathcal{N}\int\mathcal{D}\tilde{\phi} \ e^{i\mathcal{A}_{E}[\tilde{\phi},\tilde{g}]+\tilde{\mathcal{A}}[\tilde{\phi},\tilde{g}]} \nonumber\\
\tilde{\mathcal{A}}[\tilde{\phi},\tilde{g}] & =-\int\sqrt{-\tilde{g}}d^{4}x \ \log\Bigg[\cosh\left(\sqrt{\frac{4\pi G}{3}}\tilde{\phi}\right)\Bigg],
\end{align}
in the Einstein frame using the Jacobian of field transformation and taking into account the Weyl transformation of the measure defined in (\ref{measure}). Now we define the following quantity
\begin{equation}
\mathcal{T}^{\mu\nu}\equiv-\frac{2i}{\sqrt{-\tilde{g}}}\frac{\delta\tilde{\mathcal{A}}[\tilde{\phi},\tilde{g}]}{\delta\tilde{g}_{\mu\nu}}.
\end{equation}
In terms of the functional $\tilde{\mathcal{A}}[\tilde{\phi},\tilde{g}]$ and $\mathcal{T}^{\mu\nu}$, the Ward identities in the Einstein frame become
\begin{align}
i & \tilde{\nabla}_{\mu}\langle[\tilde{T}^{\mu\nu}(x)+\mathcal{T}^{\mu\nu}(x)]\prod_{i=1}^{n}\tilde{\mathcal{O}}_{i}(\tilde{\phi}(x_{i}),\tilde{g}(x_{i}))\rangle_{J-E} \nonumber\\
 & =-\frac{i}{\sqrt{-\tilde{g}(x)}}\Big\langle\tilde{\nabla}^{\nu}\tilde{\phi}(x)\frac{\delta\mathcal{A}_{2}[\tilde{\phi},\tilde{g}]}{\delta\tilde{\phi}(x)}\prod_{i=1}^{n}\tilde{\mathcal{O}}_{i}(\tilde{\phi}(x_{i}),\tilde{g}(x_{i}))\Big\rangle_{J-E} \nonumber\\
 & -\sum_{j=1}^{n}\frac{1}{\sqrt{-\tilde{g}(x)}}\Big\langle\tilde{\nabla}_{\mu}\left(\frac{\delta\tilde{\mathcal{O}}_{j}(\tilde{\phi}(x_{j}),\tilde{g}(x_{j}))}{\delta\tilde{g}_{\mu\nu}(x)}\right)\prod_{i\neq j}^{n}\tilde{\mathcal{O}}_{i}(\tilde{\phi}(x_{i}),\tilde{g}(x_{i}))\Big\rangle_{J-E} \nonumber\\
 & -\sum_{j=1}^{n}\frac{1}{\sqrt{-\tilde{g}(x)}}\Big\langle\tilde{\nabla}^{\nu}\tilde{\phi}(x)\frac{\delta\tilde{\mathcal{O}}_{j}(\tilde{\phi}(x_{j}),\tilde{g}(x_{j}))}{\delta\tilde{\phi}(x)}\prod_{i\neq j}\tilde{\mathcal{O}}_{i}(\tilde{\phi}(x_{i}),\tilde{g}(x_{i}))\Big\rangle_{J-E},
\end{align}
according to the observer in the Jordan frame, where $\mathcal{A}_{2}[\tilde{\phi},\tilde{g}]=\mathcal{A}_{E}[\tilde{\phi},\tilde{g}]-i\tilde{\mathcal{A}}[\tilde{\phi},\tilde{g}]$, and
\begin{align}
\langle & \prod_{i=1}^{n}\tilde{\mathcal{O}}_{i}(\tilde{\phi}(x_{i}),\tilde{g}(x_{i}))\rangle_{J-E}\equiv\int\mathcal{D}\tilde{\phi} \ \Bigg[\prod_{i=1}^{n}\tilde{\mathcal{O}}_{i}(\tilde{\phi}(x_{i}),\tilde{g}(x_{i})) \nonumber\\
 & \times e^{i\mathcal{A}_{E}[\tilde{\phi},\tilde{g}]-\int\sqrt{-\tilde{g}(x)}d^{4}x\log\cosh\left(\sqrt{\frac{4\pi G}{3}}\tilde{\phi}(x)\right)}\Bigg].
\end{align}
The above expressions also clearly suggest that the descriptions of a quantum field theory in the Jordan and Einstein frames are inequivalent under the frame transformations. The inequivalence between the descriptions of a quantum field theory in the Jordan and Einstein frames originates from the non-trivial Jacobian of the measure of the functional integral due to the frame transformation.

\subsection{Ward identities due to the Weyl invariance}

In this section, we consider a $D$-dimensional theory described by an action $S_{J}$ in the Jordan frame which is invariant under the Weyl transformations. The same theory is described by an action $S_{E}$ in the Einstein frame. For the mathematical simplicity, we assume this theory contains a single scalar field $\phi$. These two frames are related by the following field transformations
\begin{equation}\label{Frame transformation CFT}
\tilde{g}_{\mu\nu} = e^{2\omega(\phi)}g_{\mu\nu}, \ \phi = \mathcal{G}(\tilde{\phi})
\end{equation}
where $\omega$ and $\mathcal{G}$ depends on the theory. Let us now consider a set of observables $\{\mathcal{O}_{i}(x)\}$ in the Jordan frame which become $\{\tilde{\mathcal{O}}_{i}(x)\}$ in the Einstein frame. These observables depend on the scalar field and the metric fields in general. Since these theories in the Jordan and Einstein frames are Weyl invariant, their action remains invariant under an infinitesimal small Weyl transformation $g_{\mu\nu} \mapsto \hat{g}_{\mu\nu} = (1 + 2\Omega)g_{\mu\nu}$ and $\tilde{g}_{\mu\nu} \mapsto \hat{\tilde{g}}_{\mu\nu} = (1 + 2\Omega)\tilde{g}_{\mu\nu}$. We may note here that the correlation functions between these observables in the Jordan frame satisfy the following relations
\begin{equation}\label{CFT 1}
\begin{split}
\langle\mathcal{O}_{1}(x_{1})\ldots\mathcal{O}_{n}(x_{n})\rangle_{J} & = \mathcal{N}\int\mathcal{D}\phi \ \prod_{i = 1}^{n}\mathcal{O}_{i}(x_{i}) \ e^{iS_{J}[\phi,g]}\\
 & = \mathcal{N}\int\mathcal{D}\phi \ \prod_{i = 1}^{n}\mathcal{O}_{i}'(x_{i}) \ e^{iS_{J}[\phi,\hat{g}]}\\
 & = \mathcal{N}\int\mathcal{D}\phi \ \prod_{i = 1}^{n}\mathcal{O}_{i}'(x_{i}) \ e^{iS_{J}[\phi,g] + i\delta_{\Omega}S_{J}[\phi,g]}, 
\end{split}
\end{equation} 
where
\begin{equation}\label{CFT 2}
\begin{split}
\delta_{\Omega}S_{J} & = \int\sqrt{-g(x)}d^{D}x \ T_{ \ \mu}^{\mu}(x)\Omega(x)\\
\mathcal{O}_{i}'(x_{i}) & = \mathcal{O}_{i}(x_{i}) + 2\int d^{D}x \ \frac{\delta\mathcal{O}_{i}(x_{i})}{\delta g_{\mu\nu}(x)}g_{\mu\nu}(x)\Omega(x).
\end{split}
\end{equation}
Plugging the expressions (\ref{CFT 2}) in the equation (\ref{CFT 1}), we obtain the following Ward identities in the Jordan frame
\begin{equation}
\langle T_{ \ \mu}^{\mu}(x)\mathcal{O}_{1}(x_{1})\ldots\mathcal{O}_{n}(x_{n})\rangle_{J} = i\frac{2g_{\mu\nu}(x)}{\sqrt{-g(x)}}\sum_{j = 1}^{n}\Big\langle\frac{\delta\mathcal{O}_{j}(x_{j})}{\delta g_{\mu\nu}(x)}\prod_{i \neq j}\mathcal{O}_{i}(x_{i})\Big\rangle_{J}.
\end{equation}
Naively we might expect that the above Ward identities become the following in the Einstein frame
\begin{equation}
\langle \tilde{T}_{ \ \mu}^{\mu}(x)\tilde{\mathcal{O}}_{1}(x_{1})\ldots\tilde{\mathcal{O}}_{n}(x_{n})\rangle_{E} = i\frac{2\tilde{g}_{\mu\nu}(x)}{\sqrt{-\tilde{g}(x)}}\sum_{j = 1}^{n}\Big\langle\frac{\delta\tilde{\mathcal{O}}_{j}(x_{j})}{\delta \tilde{g}_{\mu\nu}(x)}\prod_{i \neq j}\tilde{\mathcal{O}}_{i}(x_{i})\Big\rangle_{E},
\end{equation}
where
\begin{equation}
\langle\tilde{\mathcal{O}}_{1}(x_{1})\ldots\tilde{\mathcal{O}}_{n}(x_{n})\rangle_{E} = \mathcal{N}\int\mathcal{D}\tilde{\phi} \ \prod_{i = 1}^{n}\tilde{\mathcal{O}}_{i}(x_{i}) \ e^{iS_{E}[\tilde{\phi},\tilde{g}]}.
\end{equation}
However, that is not the case from the point of view of an observer in the Jordan frame since according to this observer, we find the following relation
\begin{equation}
\mathcal{Z}_{J} = \mathcal{N}\int\mathcal{D}\phi \ e^{iS_{J}} = \mathcal{N}\int\mathcal{D}\tilde{\phi} \ e^{iS_{E} + \mathcal{A}},
\end{equation}
where 
\begin{equation}\label{correction in action}
\mathcal{A} = - \frac{(D - 2)}{2}\int\sqrt{-\tilde{g}(x)}d^{D}x \ \omega[\phi(x)] + \int\sqrt{-\tilde{g}(x)}d^{D}x \ \log\left(\frac{\delta\phi(x)}{\delta\tilde{\phi}(x)}\right).
\end{equation}
The above expression of $\mathcal{A}$ follows from the Jacobian of field transformation and the equation (\ref{measure}) in Appendix. As a result, the Ward identities in the Einstein frame \textit{w.r.t} an observer in the Jordan frame is given by
\begin{equation}
\langle \bar{T}_{ \ \mu}^{\mu}(x)\tilde{\mathcal{O}}_{1}(x_{1})\ldots\tilde{\mathcal{O}}_{n}(x_{n})\rangle_{J-E} = i\frac{2\tilde{g}_{\mu\nu}(x)}{\sqrt{-\tilde{g}(x)}}\sum_{j = 1}^{n}\Big\langle\frac{\delta\tilde{\mathcal{O}}_{j}(x_{j})}{\delta \tilde{g}_{\mu\nu}(x)}\prod_{i \neq j}\tilde{\mathcal{O}}_{i}(x_{i})\Big\rangle_{J-E},
\end{equation}
where
\begin{equation}
\begin{split}
\langle\tilde{\mathcal{O}}_{1}(x_{1}) & \ldots\tilde{\mathcal{O}}_{n}(x_{n})\rangle_{J-E} = \mathcal{N}\int\mathcal{D}\tilde{\phi} \ \prod_{i = 1}^{n}\tilde{\mathcal{O}}_{i}(x_{i}) \ e^{iS_{E}[\tilde{\phi},\tilde{g}] + \mathcal{A}[\tilde{\phi},\tilde{g}]}\\
\bar{T}_{ \ \mu}^{\mu}(x) & = -\frac{2\tilde{g}^{\mu\nu}(x)}{\sqrt{ - \tilde{g}(x)}}\frac{\delta}{\delta \tilde{g}^{\mu\nu}(x)}[S_{E} - i\mathcal{A}] = \tilde{T}_{ \ \mu}^{\mu}(x) + \frac{2i\tilde{g}^{\mu\nu}(x)}{\sqrt{ - \tilde{g}(x)}}\frac{\delta\mathcal{A}}{\delta \tilde{g}^{\mu\nu}(x)}.
\end{split}
\end{equation}
We may note here that the first term in (\ref{correction in action}) vanishes for $D = 2$. It is clear from the above expression that the descriptions of a Weyl invariant field theory in the Jordan and Einstein frames are inequivalent under the frame transformations.

\section{Effective action}
In this section, we compare the effective actions involving one-loop quantum corrections corresponding to a field theory described in two frames, namely the Jordan and Einstein frames related by the frame transformation. In order to compute these one-loop corrections, one needs to expand the action in the exponent of the generating functional around a classical solution of the equation of motion up to second order term and do the Gaussian functional integral. For simplicity, we consider the action (\ref{action 2 in diff inv}) defined in the Jordan frame. The corresponding effective action $\Gamma_{J}[\phi]$ is given by
\begin{equation}\label{eff action Jordan}
\Gamma_{J}[\phi,g]=\mathcal{A}_{J}[\phi,g]+\frac{i}{2}\log\Big[\text{det}\left(\frac{\delta^{2}\mathcal{A}_{J}}{\delta\phi(x)\delta\phi(y)}\right)\Big],
\end{equation}
where the matrix element inside the determinant is given by
\begin{equation}
\frac{\delta^{2}\mathcal{A}_{J}}{\delta\phi(x)\delta\phi(y)}=\sqrt{-g(x)}\delta^{(4)}(x-y)\left(\Box-\frac{R}{6}-U''(\phi)\right).
\end{equation}
This might suggest naively that for an observer in the Einstein frame, the effective action becomes
\begin{equation}\label{eff action Einstein}
\Gamma_{E}[\tilde{\phi},\tilde{g}]=\mathcal{A}_{E}[\tilde{\phi},\tilde{g}]+\frac{i}{2}\log\Big[\text{det}\left(\frac{\delta^{2}\mathcal{A}_{E}}{\delta\tilde{\phi}(x)\delta\tilde{\phi}(y)}\right)\Big],
\end{equation}
in a similar manner where the matrix element inside the determinant is given by
\begin{equation}
\frac{\delta^{2}\mathcal{A}_{E}}{\delta\tilde{\phi}(x)\delta\tilde{\phi}(y)}=\sqrt{-\tilde{g}(x)}\delta^{(4)}(x-y)\left(\Box_{E}-\tilde{U}''(\tilde{\phi})\right),
\end{equation}
where $\Box_{E}=\tilde{g}^{\mu\nu}\tilde{\nabla}_{\mu}\tilde{\nabla}_{\nu}$. The relation between $\tilde{U}(\tilde{\phi})$ and $U(\phi)$ is given by
\begin{equation}
\tilde{U}(\tilde{\phi})=e^{-4\omega}U(\phi), \ e^{2\omega}=1-\frac{4\pi G}{3}\phi^{2},
\end{equation}
whereas the relation between $\tilde{U}''(\tilde{\phi})$ and $U''(\phi)$ is
\begin{align}
\tilde{U}''(\tilde{\phi}) & =\Bigg[\left(\frac{\delta\phi}{\delta\tilde{\phi}}\right)^{2}\frac{\delta^{2}}{\delta\phi^{2}}+\frac{\delta^{2}\phi}{\delta\tilde{\phi}^{2}}\frac{\delta}{\delta\phi}\Bigg](e^{-4\omega}U(\phi)), \ \frac{\delta\phi}{\delta\tilde{\phi}}=e^{2\omega}.
\end{align}
However, for an observer in the Jordan frame, (\ref{eff action Einstein}) is not the effective action in the Einstein frame. In this case, we find the following effective action in the Einstein frame from the point of view of an observer in the Jordan frame
\begin{equation}\label{Final expression}
\begin{split}
\Gamma_{E-J}[\tilde{\phi},\tilde{g}] & =\mathcal{A}_{E}[\tilde{\phi},\tilde{g}]-i\tilde{\mathcal{A}}[\tilde{\phi},\tilde{g}]\\
 & +\frac{i}{2}\log\Bigg[\text{det}\left(\frac{\delta^{2}\mathcal{A}_{E}}{\delta\tilde{\phi}^{2}}+\frac{\delta^{2}\tilde{\phi}}{\delta\phi^{2}}\frac{\delta\mathcal{A}_{E}}{\delta\tilde{\phi}}\left(\frac{\delta\phi}{\delta\tilde{\phi}}\right)^{2}\right)\Bigg].
\end{split}
\end{equation}
The above result follows from the following relations
\begin{equation}\label{imp relation}
\begin{split}
\mathcal{Z}_{J} & =\mathcal{N}\int\mathcal{D}\phi \ e^{i\mathcal{A}_{J}[\phi,g]}=\mathcal{N}\int\mathcal{D}\tilde{\phi}\sqrt{\text{det}\left(\frac{\delta\phi}{\delta\tilde{\phi}}\right)} \ e^{i\mathcal{A}_{J}[\phi,g]}\\
 & \approx\mathcal{N}\int\mathcal{D}\tilde{\phi}\sqrt{\text{det}\left(\frac{\delta\phi}{\delta\tilde{\phi}}\right)_{\phi_{c}}} \ e^{i\mathcal{A}_{J}[\phi_{c},g]+\frac{i}{2}\eta .\frac{\delta^{2}\mathcal{A}_{J}[\phi_{c},g]}{\delta\phi^{2}}.\eta},
\end{split}
\end{equation}
where we expanded $\mathcal{A}_{J}$ around the solution of equation of motion by considering $\phi=\phi_{c}+\eta$, and
\begin{equation}
\eta .\frac{\delta^{2}\mathcal{A}_{J}[\phi_{c},g]}{\delta\phi^{2}}.\eta\equiv\int d^{4}x d^{4}y \ \eta(x)\frac{\delta^{2}\mathcal{A}_{J}[\phi_{c},g]}{\delta\phi(x)\delta\phi(y)}\eta(y).
\end{equation}
In (\ref{imp relation}), both the Jacobian of field transformation and the change in measure under the conformal transformation are taken into account. The corresponding $\tilde{\phi}$ is given by $\tilde{\phi}=\tilde{\phi}_{c}+\tilde{\eta}$ where $\eta=\frac{\delta\phi}{\delta\tilde{\phi}}\tilde{\eta}$, and using it, we obtain the following relation
\begin{equation}
\begin{split}
\mathcal{Z}_{J} & \approx\mathcal{N}\int\mathcal{D}\tilde{\eta}\sqrt{\text{det}\left(\frac{\delta\phi}{\delta\tilde{\phi}}\right)_{\phi_{c}}} \ e^{i\mathcal{A}_{E}[\tilde{\phi}_{c},\tilde{g}]+\frac{i}{2}\tilde{\eta}\frac{\delta\phi}{\delta\tilde{\phi}}.\frac{\delta^{2}\mathcal{A}_{J}[\phi_{c},g]}{\delta\phi^{2}}.\frac{\delta\phi}{\delta\tilde{\phi}}\tilde{\eta}}, \  \sqrt{\text{det}\left(\frac{\delta\phi}{\delta\tilde{\phi}}\right)} = e^{\int\sqrt{-g(x)}\omega(x)d^{4}x}.
\end{split}
\end{equation}
After doing the above Gaussian integral and taking the logarithm of both sides, we obtain the result (\ref{Final expression}) using the definition $i\Gamma=\log\mathcal{Z}$. Thus, from the above-mentioned expressions of effective actions, we are also able to establish that these two frames are inequivalent at the quantum level. The differences in these expressions of effective actions come from the Jacobian of the frame transformation. In \cite{herrero2016anomalies}, the second term in the equation (\ref{Final expression}) is not considered while deriving the expressions for effective action. The terms containing the logarithm of determinants can be computed in principle using the functional trace in the heat-kernel method \cite{vassilevich2003heat, avramidi1990nonlocal, avramidi1991covariant}.

\section{Discussion}
The frame transformation from the Jordan frame to the Einstein frame is done often in order to reduce the computational complexity. Further, in order to test the scalar-tensor theories of gravity, observables computed in the Einstein frame are often used for comparison with experimental observations in the Jordan frame \cite{faulkner2007constraining, sotani2017maximum, ooba2016planck, scharer2014testing, barausse2013neutron}. Therefore, unless these frames are equivalent to each other exactly, such comparisons with astrophysical and cosmological observations are always questionable. In this article, it is shown that in a generic scalar-tensor theory of gravity, the physical descriptions in the Jordan and Einstein frames are not equivalent despite being related through Weyl transformation. This is shown explicitly by comparing the Ward identities in quantum field theory according to an observer in the Jordan frame and another observer in the Einstein frame, connected via frame transformation. Further, the equivalence between these two frames is broken at the quantum level due to the presence of additional non-vanishing terms in the Ward identities which is shown explicitly. We conclude the same by comparing the expressions of effective actions in a quantum field theory for the above-mentioned observers connected via frame transformation. The above inequivalence originates from the change in the quantum state under the frame transformation. In the present article, it is shown that the vacuum state in one frame does not coincide with the vacuum state in the other frame under a frame transformation, which follows from the difference in the functional integrals computed in the above two frames. In \cite{falls2019frame}, the inequivalence between the Jordan and Einstein frames is also discussed. The approach considered in \cite{falls2019frame} is different from ours, although the main idea is almost the same. Though our approach is restricted within quantum field theories in a classical spacetime in which the metric components are not treated as dynamical degrees of freedom, it is simpler in showing the inequivalence between the observers connected via a frame transformation. 

Although our approach outlined in the present article is semi-classical, our result is robust. It will also be valid for a full quantum theory of gravity, and the reason is as follows. Firstly, in any functional integral formulation of quantum gravity, one must also consider the functional integral over metric field variables. Therefore, under the frame transformation, we expect that there will be a non-trivial Jacobian of the frame transformation. Further, the approach that we considered semi-classically in the present article will also be present there. As a result, this effect cannot be avoided under any circumstances. As a result, Einstein and Jordan frames remain inequivalent at the full quantum theory of gravity under frame transformations. This has particular relevance in quantum cosmology, where observables for non-minimally coupled field theories are often computed in the Einstein frame through suitable frame transformations for mathematical convenience.
 
\section{Acknowledgements}
The author thanks SERB grant (Project RD/0122-SERB000-044) for funding this research.

\bibliographystyle{unsrt}
\bibliography{draft}

\appendix

\section{Functional integral formulation}

In this appendix, we derive the measure used in the functional integral formulation of quantum field theory in curved spacetime. Let us consider a generic scalar field theory, described by the action of the following form
\begin{align}
\mathcal{A} =-\int\sqrt{-g(x)}d^{4}x\Big[\frac{1}{2}g^{\mu\nu}(x)D^{ab}(x)\partial_{\mu}\phi_{a}(x)\partial_{\nu}\phi_{b}(x)+U[\{\phi_{a}(x)\}]\Big].
\end{align} 
The corresponding canonical momentum conjugate to the field variable is given by
\begin{equation}
\begin{split}
\Pi^{a}(x) & =\frac{\partial\mathcal{L}}{\partial(\partial_{0}\phi_{a}(x))}=-g^{0\nu}(x)\sqrt{-g(x)}D^{ab}(x)\partial_{\nu}\phi_{b}(x),
\end{split}
\end{equation}
and the above equation can be inverted to write the following relation
\begin{equation}
\partial_{0}\phi_{a}(x)=-\frac{1}{g^{00}(x)}\Big[\frac{D_{ab}(x)\Pi^{b}(x)}{\sqrt{-g(x)}}+g^{0i}(x)\partial_{i}\phi_{a}(x)\Big],
\end{equation}
where $D_{ab}(x)\equiv(D^{-1}(x))^{ab}$. As a result, the corresponding Hamiltonian density is given by
\begin{equation}
\begin{split}
 & \mathcal{H}(x)=\Pi^{a}(x)\partial_{0}\phi_{a}(x)-\mathcal{L}\\
 & =-\frac{1}{g^{00}(x)}\Big[\frac{D_{ab}(x)}{\sqrt{-g(x)}}\Pi^{a}(x)\Pi^{b}(x)+g^{0i}(x)\Pi^{a}(x)\partial_{i}\phi_{a}(x)\Big]\\
 & +\sqrt{-g(x)}\Big[\frac{1}{2}g^{\mu\nu}(x)D^{ab}(x)\partial_{\mu}\phi_{a}(x)\partial_{\nu}\phi_{b}(x)+U[\{\phi_{a}(x)\}]\Big]\\
 & =-\frac{1}{2g^{00}(x)}\Big[\frac{D_{ab}(x)}{\sqrt{-g(x)}}\Pi^{b}(x)+2g^{0i}(x)\partial_{i}\phi_{a}(x)\Big]\Pi^{a}(x)\\
 & +\sqrt{-g(x)}\Big[\frac{1}{2}g^{ij}(x)D^{ab}(x)\partial_{i}\phi_{a}(x)\partial_{j}\phi_{b}(x)+U[\{\phi_{a}(x)\}]\Big]\\
 & -\sqrt{-g(x)}\frac{1}{2g^{00}(x)}D^{ab}(x)g^{0i}(x)\partial_{i}\phi_{a}(x)g^{0j}(x)\partial_{j}\phi_{b}(x).
\end{split}
\end{equation}
Hence, the generating functional or the partition functional is given by
\begin{equation}\label{measure}
\begin{split}
\mathcal{Z} & =\int\mathcal{D}\Pi^{a}\mathcal{D}\phi_{a} \ e^{i\int d^{4}x[\Pi^{a}(x)\partial_{0}\phi_{a}(x)-\mathcal{H}[\Pi^{a}(x),\phi_{a}(x)]]}\\
 & =\int\prod_{a}\left(\mathcal{D}\phi_{a}\Big[\prod_{x}\text{det}[D^{cd}(x)]|g^{00}(x)|\sqrt{-g(x)}\Big]^{\frac{1}{2}}\right)e^{i\mathcal{A}[\{\phi_{a}\}]}.
\end{split}
\end{equation}
In order to see the appearance of the action in the exponent on the last line, we express the quantity $\Pi^{a}(x)\partial_{0}\phi_{a}(x)-\mathcal{H}(x)$ as
\begin{equation}
\begin{split}
\Pi^{a}(x)\partial_{0}\phi_{a}(x) - \mathcal{H}(x) & =-\frac{1}{2|g^{00}(x)|\sqrt{-g(x)}}\Big[D_{ab}(x)\Pi^{a}(x)\Pi^{b}(x)+2\sqrt{-g(x)}\Pi^{a}(x)\partial^{0}\phi_{a}(x)\Big]\\
 & -\sqrt{-g(x)}\Big[\frac{1}{2}g^{ij}(x)D^{ab}(x)\partial_{i}\phi_{a}(x)\partial_{j}\phi_{b}(x)+U[\{\phi_{a}(x)\}]\Big]\\
 & +\sqrt{-g(x)}\frac{1}{2g^{00}(x)}D^{ab}(x)g^{0i}(x)\partial_{i}\phi_{a}(x)g^{0j}(x)\partial_{j}\phi_{b}(x).
\end{split}
\end{equation}
Now doing the functional integral over the conjugate momentum variables $\Pi^{a}(x)$, we obtain the four-dimensional spacetime integral over the following quantity in the exponent
\begin{align}
-\frac{1}{2g^{00}(x)} & \sqrt{-g(x)}D^{ab}(x)\partial^{0}\phi_{a}(x)\partial^{0}\phi_{b}(x)\\
-\sqrt{-g(x)} & \Big[\frac{1}{2}g^{ij}(x)D^{ab}(x)\partial_{i}\phi_{a}(x)\partial_{j}\phi_{b}(x)+U[\{\phi_{a}(x)\}]\Big] \nonumber\\
+\sqrt{-g(x)} & \frac{1}{2g^{00}(x)}D^{ab}(x)g^{0i}(x)\partial_{i}\phi_{a}(x)g^{0j}(x)\partial_{j}\phi_{b}(x), \nonumber\\
=-\frac{1}{2}\sqrt{-g(x)} & D^{ab}(x)\Big[g^{00}(x)\partial_{0}\phi_{b}(x)+2g^{0i}(x)\partial_{i}\phi_{b}(x)\Big]\partial_{0}\phi_{a}(x) \nonumber\\
-\sqrt{-g(x)} & \Big[\frac{1}{2}g^{ij}(x)D^{ab}(x)\partial_{i}\phi_{a}(x)\partial_{j}\phi_{b}(x)+U[\{\phi_{a}(x)\}]\Big] \nonumber\\
=-\sqrt{-g(x)} & \Big[\frac{1}{2}g^{\mu\nu}(x)D^{ab}(x)\partial_{\mu}\phi_{a}(x)\partial_{\nu}\phi_{b}(x)+U[\{\phi_{a}(x)\}]\Big], \nonumber
\end{align} 
by lowering the indices of partial derivatives. According to our convention of the metric signature, $g^{00}$ is negative-definite which we used in showing the above result.

If $D^{cd}(x)$ and the metric are independent of field variables, then the measure $\Big[\prod_{x}\text{det}[D^{cd}(x)]g^{00}(x)\sqrt{-g(x)}\Big]^{\frac{1}{2}}$ can be dropped from the functional integral as it is dynamically irrelevant. For more details, see \cite{coleman1988aspects, de1983functional, toms1987functional}. However, for a field-dependent Weyl transformation in the frame transformations, the change in the above-mentioned measure must be taken into account.

\end{document}